\documentclass[useAMS,usenatbib]{mn2e}
\usepackage{epsfig,graphics,graphicx,bm,amssymb}

\begin{document}

\title{Dynamical Stability of Terrestrial Planets in the Binary $\alpha$ Centauri System}

\author[E. Andrade-Ines and T.A. Michtchenko]
{E. Andrade-Ines\thanks{E-mail: eduardo.andradeines@iag.usp.br} and T.A. Michtchenko\\
\\
Instituto de Astronomia, Geof\'{\i}sica e Ci\^encias Atmosf\'ericas, USP, Rua do Mat\~ao 1226, 05508-090 S\~ao Paulo, Brazil }

\maketitle

\begin{abstract}

In this paper, we investigate whether hypothetical Earth-like planets have high probability of remaining on stable orbits inside the habitable zones around the stars A and B of $\alpha$ Centauri, for lengths of time compatible with the evolution of life. We introduce a stability criterion based on the solution of the restricted three-body problem and apply it to the $\alpha$ Centauri system. In this way, we determine the regions of the short-term stability of the satellite-type (S-type) planetary orbits,  in both planar and three-dimensional cases. We also study the long-term stability of hypothetical planets through the dynamical mapping of the habitable zones of the stars. The topology of the maps is analyzed using the semi-analytical secular Hamiltonian model and possible processes responsible for long-lasting instabilities are identified.  We verify that the planetary motion inside the habitable zones is regular, regardless of high eccentricities, for inclinations smaller than $40^\circ$. We show that the variation of the orbital distance of the planet located in the habitable zones of the binary is comparable to that of Earth, if the planet is close to the Mode I stationary solution. This result brings positive expectations for finding habitable planets in binary stars.
\end{abstract}


\section{Introduction}

About 50\% of the known main-sequence-stars belong to the class of binary or multiple stellar systems \citep{1979AJ.....84.1591A,1991A&A...248..485D,2010ApJS..190....1R}. Due to inherent difficulties in monitoring the radial velocities of multi-star systems,  these systems have been mostly ignored by researchers performing searches for extra-solar planets.  However, in last years, due to improving precision and advances of detection methods, binaries have become increasingly interesting targets of planet searches, and several research programs looking for planets in binary systems are in progress.

The $\alpha$ Centauri binary is of special interest. Being the closest stellar system to the Sun, as well as composed of two stars very similar to the Sun, it has been always a major target for extrasolar planet surveys, although it was not until recently that a putative planet has been found orbiting around $\alpha$ Centauri B with a period of 3.236 days (\citealp{2012Natur.491..207D}; for an opposing view see \citealp{2013ApJ...770..133H}). Furthermore, studies of the planet formation suggest that more planets around the star B may exist, particularly inside the \emph{Habitable Zone} (HZ) \citep{2008ApJ...679.1582G,2009MNRAS.393L..21T}. Although these planets are yet to be discovered in binary systems, recent works have shown that terrestrial planets can form and have a stable orbit around a star of a binary in a close S-type system, depending on the binary orbital and physical parameters \citep{2006ApJ...644..543H,2007ApJ...666..436H,2007ApJ...660..807Q,2013MNRAS.428.3104E,2013ApJ...777..165K}.

The high mass of the disturbing stellar companion and the frequent high eccentricity/inclination of its relative orbit make the perturbation theories based on classical Lagrangian expansions inapplicable in this case. A possible approximation is the expansion of the disturbing function using the Legendre polynomials \citep{1978A&A....65..421H,2000ApJ...535..385F,2010A&A...522A..60L}. However, although this development avoids the constraints due to the high eccentricities and inclinations, its convergence is very slow, allowing its application only in the case of hierarchical systems \citep{2000ApJ...537L..65N,2006ApJ...641.1148B,2006ApJ...644..543H,2009MNRAS.393L..21T,2013MNRAS.428.3104E}.

An alternative to overcome the problem is the extension of a perturbation theory up to second order in the disturber  mass. This was done in \cite{2011A&A...530A.103G}, where the developed model was  applied to the $\gamma$ Cephei binary system. Although the results obtained have shown an improvement over the first-order model, the authors have emphasized that the model was not workable due to its high complexity; thus they preferred the use of empiric correction terms, specific for the $\gamma$ Cephei configuration.

In this work, we perform a study of the stability of the planetary motion in binary systems combining the analytical and numerical approaches. For this, the stability concept is separated into the short-term stability and the long-term stability.  The short-term stability is based on the Hill criterion, which does not require any kind of expansion and averaging of the disturbing function. In such a way, it does not introduce the constrains on neither the mass of the disturbing body nor the eccentricity/inclination of its orbit.

The Hill criterion is a powerful approach, which has been used for over a century \citep{1905Hill}; an in-depth review of its applications can be found in \cite{1984CeMec..34...49S}. Analytical approaches based on the Hill stability concept have been developed in \cite{1982CeMec..26..311M}, \cite{1993Icar..106..247G}, among others, also as some empirical relations between stable orbits and the physical parameters of the binary \citep{1981ApJ...251..337G,1982AJ.....87.1333B,1983AJ.....88.1415P,1988A&A...191..385R,1999AJ....117..621H}. In our paper, we develop a variation of the Hill approach, referring to it as Maximum Apocentric Distance (or MAcD--criterion), and apply it to  the $\alpha$ Centauri stars A and B. To simplify the applications of the method, we present the explicit formula for the maximal apocentric distances as a function of the physical and orbital parameters of the stellar binary.

The study of the long-term stability of a binary-planet system (of the order of the lifetime of the stars) is done for the S-type (satellite type) planetary motion, when the planet revolves around one of the stellar components \citep{1986BAAS...18..842D}. The planet motion is modeled semi-analytically in the frame of the secular three-body problem \citep{2006Icar..181..555M} and applied to the $\alpha$ Centauri hypothetical Earth-like planets placed in the HZs of the stars A and B. The topology of the phase space is investigated, without any restriction on the magnitude of the eccentricities and inclinations of the planets. The boundaries of the secular stability of the system, as a function of the masses and the orbital parameters of the planets, are then obtained. Notwithstanding the averaging being done at first order in masses, the relatively large separation between the stars allows a good agreement of the analytical results with those obtained numerically. The direct numerical investigations are done through the dynamical mapping of the regions of habitability of both stars in the $\alpha$ Centauri system.

The physical and orbital parameters of the $\alpha$ Centaury binary used in this paper were taken from \cite{2002A&A...386..280P}. They are the masses of the components A and B, $m_A = 1.105 \pm 0.0070 M_\odot$ and $m_B = 0.934 \pm 0.0061 M_\odot$, respectively, and the eccentricity $e=0.518 \pm 0.0013$  and the semimajor axis $a = 17.57 \pm 0.022$ $arcsec$ (or $a=23.4 \pm 0.03$\,AU, if the parallax of $746.8 \pm 1.2$ $mas$ is adopted) of the relative orbit.

This paper is structured as in the following. In the next section, we study the short-term stability of the system, introducing a stability criterion and applying it to both the planar and 3D cases for planets on S-type orbits. In Section \ref{sec2} we study the long-term stability, applying the semi-analytical 3D model and comparing the results to those obtained through dynamical mapping. In Section \ref{sec3}, we investigate whether the planets remain inside the HZ around the central star, regardless of large amplitudes of short-term oscillations provoked by the stellar companion. Finally, the summary is given in Section \ref{sum}.


\section{Short-Term Dynamical Stability}\label{sec1}

In this section, we introduce the stability criterion for S-type planetary orbits in binary star systems based on the concept of the Hill stability \citep[ among others]{1982CeMec..26..311M,1993Icar..106..247G} and apply it to both the planar and 3D motions of the fictitious planets in the  $\alpha$ Centaury system.

\subsection{Planar model}\label{sec1-1}

Our approach is based on the classical solution of the restricted three-body problem and, in particular, on the Zero Velocity Surface method.  Zero velocity surfaces are levels of the $S(\vec r)$--function defined by the position of the planet with respect to the two stars \citep{1982CeMec..26..311M}. The position vectors are defined with respect to the center of mass of the system in the non-inertial rotating reference frame, in which the two stars always lie along the $x$--axis.

\begin{figure}
\centering
\includegraphics[width=1.0\columnwidth]{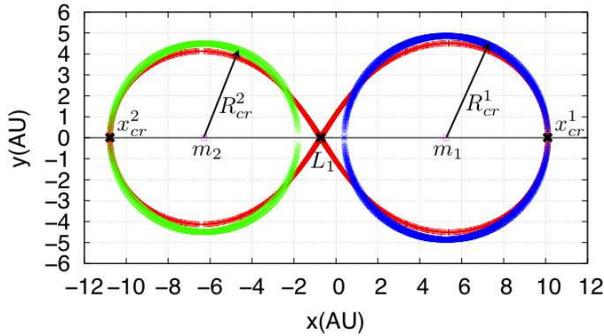}
\caption{Zero velocity curve corresponding to the Lagrangian $L_1$ solution (red curve), obtained for the stars at the pericenter of their relative orbit.  The circumferences of stability, of radii $R^1_{cr}$ around $m_1$, and $R^2_{cr}$ around $m_2$, are shown by blue and green curves, respectively.}
\label{zerovelAB}
\end{figure}

Let us consider a system composed by two stars, 1 and 2, and a planet $p$. The $S(\vec r)$-- function, also known as Jacobi integral or relative energy, can be written as

\begin{equation}
S(\vec r_p) =\sum_{i}\displaystyle\frac{m_i}{M} \Bigl[ \frac{(\vec r_p - \vec r_i)^2}{d^2} + \frac{2 d}{|\vec r_p - \vec r_i|} \Bigr]  ,
\label{super1}
\end{equation}
where $m_i$ and $\vec r_i$ ($i=1,2$) are the stellar masses and the position vectors  and $\vec r_p$ is the position vector of the planet; $M$ is the total mass of the system and $d = |\vec r_1 - \vec r_2|$. Figure \ref{zerovelAB} shows the positions of the stars $m_1$ and $m_2$ separated by the distance $d$. In the rotating frame, the function $S(\vec r_p)$ allows five equilibrium solutions (three co-linear and two triangular Lagrangian points), all located in the plane of the motion of the stars, which will be chosen as a reference plane. For our purpose, the co-linear solutions, and, particularly, $L_1$ solution, are of special interest. These solutions are located on the $x$--axis and their co-ordinates can be obtained searching for singular points on the surface (\ref{super1}), where the vectors $\vec r$ are just replaced by scalar $x$--component of the corresponding body.

It should be noted that the expression (\ref{super1}) was developed in the uniformly rotating reference frame, when two stars move on circular orbits around the baricenter of the system and their mutual distance $d$ is unchanged. To apply it to the case when the stars have elliptic orbits, it is reasonable to adopt a minimal value for $d$, when the stars are at the pericenter of their relative orbit. In this configuration, the perturbations on the S-like planetary orbit are maximal and possible instabilities may appear.

Figure \ref{zerovelAB} shows the projection of one zero-velocity surface on the reference plane (red curve); the surface was calculated using the orbital elements of the $\alpha$ Centauri binary and the minimal distance between the stars. This level holds the Lagrangian solution $L_1$, which appears as a singular saddle-like point on the $x$--axis. The same level intersects the $x$--axis at two other points; the positions of these points, $x^i_{cr}$, can be calculated from the condition

\begin{equation}
S(x^i_{cr}) - S(x_{L_1}) = 0, \hspace{1cm} i=1,2. \label{super3}
\end{equation}

We assume that these intersections define a sphere of stability around each star. By definition, the sphere of stability is centered on the corresponding star and its radius is given by $R^i_{{cr}} = |x^i_{cr} - x_i|$, where $x^i_{cr}$ is the abscissa of the  closest to the star intersection, while the abscissa of the star is $x_i$. The projections of the spheres of stability on the reference plane are shown in Figure \ref{zerovelAB}, as a blue circumference centered on the star 1 and a green circumference centered on the star 2.

It is worth noting that Equation (\ref{super3}), which is just a polynomial function of $x^i_{cr}$,  can be easily solved numerically. For the sake of simplicity, we present here approximate expressions that allow us to obtain the radii. The expressions are obtained applying the method of least squares. Without loss of generality, we assume that the star 1 is a central star, while the star 2 is a perturber.  In the case when the stellar mass ratio $\mu=m_2/m_1$ belongs to the interval $0.1 \leq \mu \leq 1.0$, the radius of the sphere of stability is given as

\begin{equation}
\begin{array}{l}
R_{cr}/a\approx  0.66823 - 0.63740e - 0.74549 \mu+ \\
+0.45496e \mu + 1.0492 \mu^2 -0.23179 e \mu^2 + \\
- 0.87722 \mu^3 + 0.31541 \mu^4, \\
\end{array}
\label{approximation1}
\end{equation}

\noindent where $a$ and $e$ are the semimajor axis and the eccentricity of the stellar relative orbit, respectively. For the stellar mass ratio $\mu=m_2/m_1$ from the interval $1.0 < \mu \leq 10.0$, the radius of the sphere is given as

\begin{equation}
\begin{array}{l}
R_{cr}/a\approx 0.45265 - 0.41921e - 0.070754\mu + \\
+0.039617e \mu + 0.010865\mu^2 - 2.1394\times10^{-3}e \mu^2 + \\
-9.3729\times10^{-4}\mu^3  + 3.3886\times10^{-5}\mu^4.\\
\end{array}
\label{approximation2}
\end{equation}

Both approximations have errors smaller than $10\%$ of the exact solutions, for $0\leq e \leq 0.9$. The error drops to less than $4\%$, considering the interval of eccentricities $0\leq e \leq 0.7$.

Now we introduce our criterion of stability. It states that the stable motion of the planet orbiting one of the stars must satisfy, at any instant, the following condition
\begin{equation}
a_p(1+e_p) < R_{{cr}}, \label{SPSeq1}
\end{equation}
where $a_p$ and $e_p$ are the planetary  osculating semimajor axis and eccentricity, with respect to the corresponding central star, and $R_{cr}$ is the radius of the sphere of stability of this star. The interpretation of the expression (\ref{SPSeq1}) is simple: since the orbital evolution of the planet is confined to the inside of the circumference of radius $R_{cr}$, even at the most distant position from the central star (at apocenter), the stability of this motion is guaranteed by the deficit of the relative energy (\ref{super1}) needed to cross the zero-velocity level holding the $L_1$ solution. It is worth noting that, due to the restricted three-body approximation, the size of the domains of stability is defined by solely the physical and orbital parameters of the stars.

Hereafter we will refer to our criterion as \textit{Maximum Apocentric Distance} or MAcD--criterion.  It assures that planetary orbits that exceed the critical distance with respect to the central star (equivalent to the radius of the sphere of stability of this star) will be unstable. However, it is worth stressing that this condition does not guarantee the long-term (of order of the lifetime of the stars) stability of the planet motion. The secular dynamics of the system and its stability will be discussed in Section \ref{sec2}.

\begin{figure}
\centering
\includegraphics[width=1.0\columnwidth]{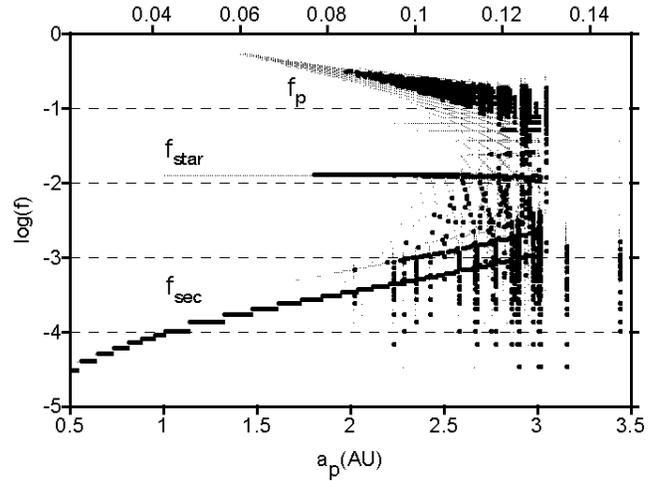}
\caption{Dynamical Power Spectrum: the proper oscillation frequencies as functions of the semimajor axis of the planet orbiting the star $\alpha$ Centauri A, in logarithmic scale. The top axis is the planetary semimajor axis in unit of the semimajor axis of the binary relative orbit. The frequency unit is $1/yr$.
}
\label{dinamespec}
\end{figure}

\subsection{Application to the $\alpha$ Centauri AB System}

In this section, we apply the  MAcD--criterion to analyze the stability of planetary S-type orbits in the $\alpha$ Centauri AB binary.

The first step is to determine the radii of the spheres of stability of the stars A and B. In the case when the star A is a central star and the star B is a perturber, the stellar mass ratio is $\mu = m_B/m_A = 0.849 < 1.0$ and the expression (\ref{approximation1}) must be used. In the case when the central star is B and the perturbing star is A, $\mu = m_A/m_B = 1.178 > 1.0$ and the expression (\ref{approximation2}) is applied. As a result, for given values of the semimajor axis and eccentricity of the relative stellar orbit, $a = 23.4$\,AU and $e = 0.518$, we obtain the critical distances for the stars A and B as $R^A_{cr} = 4.7$\,AU and $R^B_{cr} =  4.4$\,AU, respectively. The spheres of stability of the stars A and B are shown in Figure \ref{zerovelAB}, where $m_1 = m_A$ and $m_2=m_B$.


In the following step, we numerically integrate the system, in order to determine the stability of the planetary orbit defined by the condition (\ref{SPSeq1}). In this stage, the choice of the suitable time-length is important: it should be neither too large to raise the computational costs, nor too small to lose information on the proper oscillations in the planetary motion. These oscillations are usually separated in the short-period terms, associated with the Keplerian motions of the planet and the perturbing star, and the long-period secular terms, associated with the precession of the planetary orbit. Eventually, depending on the ratio of the mean motions of the planet and the perturbing star, there are also mean-motion resonance terms whose periods are intermediate lying between the short and the long periods; the resonant conditions will not be studied in this paper.

Although the MAcD-criterion is applied to assess the short-term stability of the planet motion, the long-period secular terms are still important. This is due to the fact that, during the precession, the pericenter line of the planetary orbit becomes, at some instant, aligned with the pericenter line of the orbit of the binary. If, at this instant, the perturbing star is at the pericenter of its relative orbit, the short-period perturbations become more intense and instabilities may appear. Thus, the choice of the integration timespan must account for the proper secular period of the planetary motion.

To determine the secular period, we apply the \emph{Dynamical Power Spectrum} method \citep{2002Icar..158..343M,2005LNP...683..219F}. This method is based on the Fourier decomposition of oscillations of the orbital elements and provides an information on the proper frequencies of the system under study and their evolution with changing parameters. In our study, we have chosen the semimajor axis of the planet as a parameter, which varied in the range between 0.5\,AU and 3.5\,AU. The proper frequencies of the S-type planet orbiting $\alpha$ Centauri A obtained for each value of planetary semimajor axis over $10^5$ years, are presented in Figure \ref{dinamespec}. We can clearly observe the presence of the three distinct frequency bands associated to three degrees-of-freedom of the problem. The first band, located in the high-frequency interval  $-2.0 < \log(f) < 0$, is composed of the orbital frequency of the planet, $f_p$, and its linear combinations with other proper frequencies. According to the third Kepler's law, this frequency decreases with   decreasing $a_p$. The second band is confined to the very close neighborhood of $\log(f)=-1.9$ and corresponds to the orbital frequency of the binary relative orbit, $f_{star}$.

The third band is located in the low-frequency interval $-5<\log(f)<-3$ and corresponds to the secular frequency, $f_\Delta\varpi$. This frequency tends to increase with increasing $a_p$, due to increasing perturbations on the planetary orbit from the stellar companion. The overlap of the bands occurs at $a_p > 2.5$\,AU, introducing the long-term instabilities in the planetary motion. The dense vertical lines visible already at $a_p > 2.0$\,AU in Figure \ref{dinamespec}, are associated with the mean-motion resonances in this region. Finally, for $a_p > 3.0$\,AU, mostly all orbits are disrupted during $10^5$ years.

From Figure \ref{dinamespec}, we obtain that the secular period  of the planet in the $\alpha$ Centauri system varies approximately from  $10^3$\,years up to $10^4$\,years. Assuming the integration timespan within this range, we can now numerically integrate the planetary path, in order to verify the stability of the system defined by the condition (\ref{SPSeq1}).

The quality of the prediction obtained with the MAcD-criterion can be verified by constructing dynamical maps around each star of the $\alpha$ Centauri binary. For this, we use a grid of $384 \times 100$ initial conditions on the ($a_p, e_p$) representative plane and a terrestrial planet of mass $m_p = 5\times10^{-5}M_\odot$. Without loss of generality, we fix the initial values of the mean longitudes at $\lambda_A = \lambda_B = 0^\circ$ and of the longitudes of pericenter at $\Delta\varpi=\varpi_{A,B}-\varpi_p = 0^\circ$ and $180^\circ$.  Each initial condition is numerically integrated using two different timespans: one is equal to $10^3$ years ($\sim 12.5$ orbital periods of the binary) and the second is equal to $10^4$ years ($\sim 125$ orbital periods of the binary), which cover, respectively, $\sim 1$ and $\sim 10$ secular periods at $a_p \approx 3$\,AU.

\begin{figure*}
\centering
\includegraphics[width=0.98\textwidth]{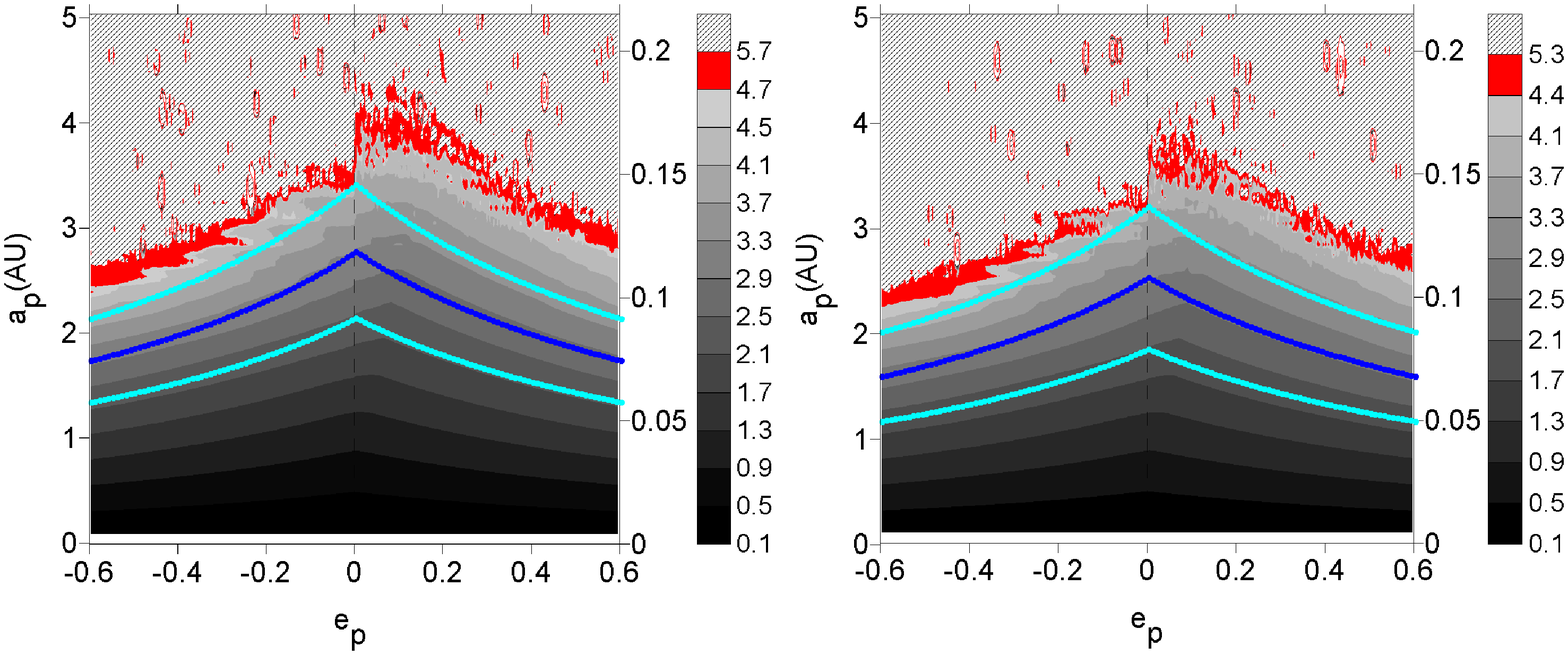}
\includegraphics[width=0.98\textwidth]{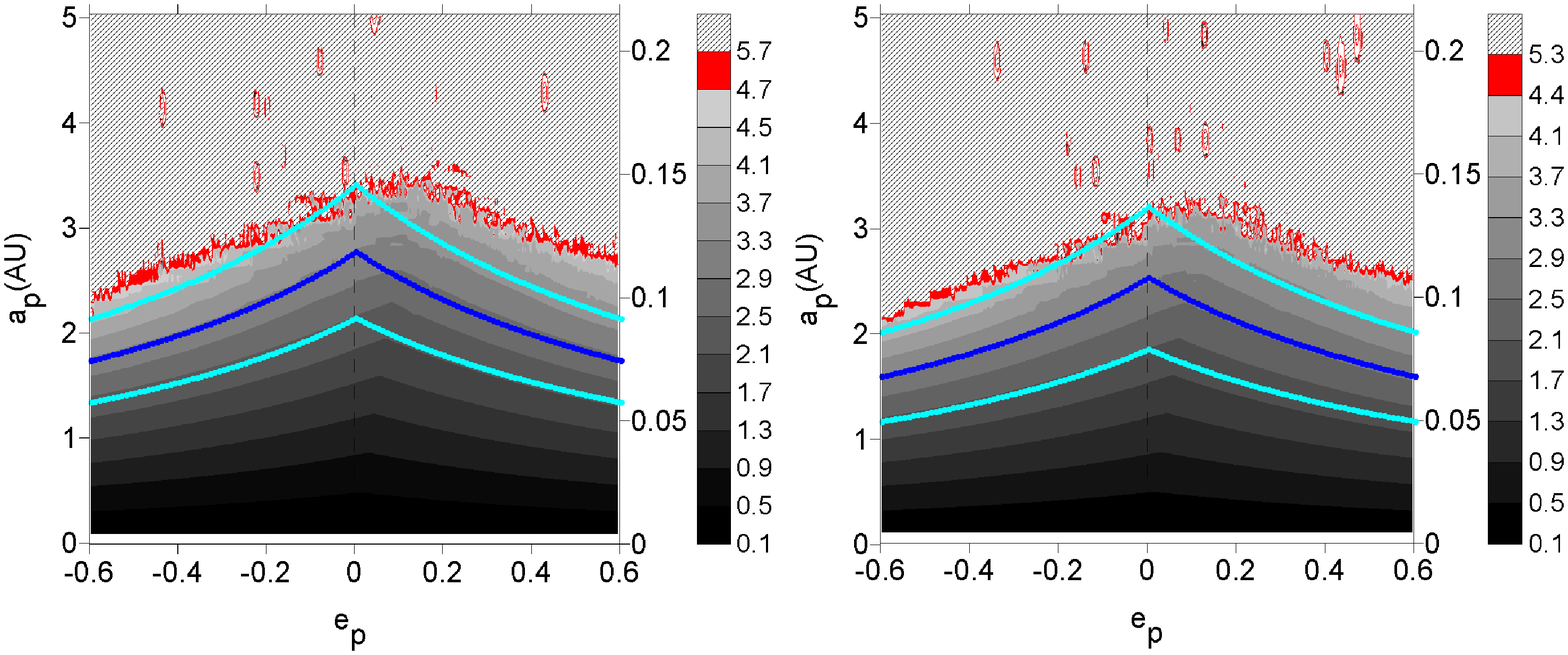}
\caption{Dynamic maps around $\alpha$ Centauri A (left panels) and B (right panels): short-term stability in the planar case. The grey scale code represents the maximal (apocentric) distance of the planet from the central star during $10^3$\,yr (top panels) and $10^4$\,yr (bottom panels) of evolution; the bars relate the grey tones to the values of the apocentric distances (in astronomical units). The positive and negative values along the $x$-axis indicate that the initial values of $\Delta \varpi$ are $0^\circ$ or $180^\circ$, respectively. The hatched regions correspond to the ejected from the system planets. The red curves correspond to levels of the radii of the stability spheres ($R^A_{cr} = 4.7$\,AU and $R^B_{cr} =  4.4$\,AU). The blue curves show the critical distances given by the Holman \& Wiegert (1999) criterion of stability, while the cyan curves show the boundaries of the corresponding grey areas (see text for details). }
\label{zerovel-all}
\end{figure*}

The solutions obtained are then analyzed applying the MAcD-criterion, in order to calculate the maximal apocentric distances that the planet reaches over the assumed timespan. The values obtained are presented using the grey scale code in Figure \ref{zerovel-all}, where the domains of the strongly unstable motion characterized by collisions and escape of the planet are dashed. On the planes we also plot, by the red curves, the levels whose values correspond to the radii of the spheres of stability, obtained using Equations \ref{approximation1} and \ref{approximation2} as  $4.7$\,AU, for the star A, and $4.4$\,AU, for the star B.

As expected, the apocentric distances increase with the increasing planetary semimajor axis and decay with the increasing eccentricity.  In the domains of the regular motion, the levels are smooth functions of the orbital elements of the planet. This behaviour is changed when the initial conditions approach the domains of the unstable motion, whose boundaries are defined, according to our MAcD-criterion, by the critical distances shown by the red curves in Figure \ref{zerovel-all}. In this case, the critical levels show erratic behaviour, characteristic of chaotic orbits. It is worth noting the good agreement between the positions of the critical levels (red curves) and the boundaries of the domains of instabilities provided by purely numerical simulations. This confirms the quality of our criterion predictions. Moreover, the comparison of the topology of the dynamical maps obtained over 1 secular period of the planets (top row in Figure \ref{zerovel-all}) and those obtained over 10 secular periods (bottom row), shows no significant differences. This means that the time-length of 1 secular period is a good choice for the application of the MAcD-criterion.

\subsection{Comparison with the Holman \& Wiegert (1999) criterion}

A similar criterion was proposed in \cite{1999AJ....117..621H}: it states that, for planetary motion around one of the stars to be stable, the initial value of the semimajor axis of the planet must obey a relationship $a_p < F_{cr}(a,e, \mu)$, where $F_{cr}$ is a function of the physical and orbital characteristics of the stars.   By means of a series of numerical integrations, the authors have found an empirical expression for $F_{cr}$ as

\begin{equation}
\begin{array}{l}
F_{cr}(a,e)/a = (0.464 \pm 0.006)+(-0.380 \pm 0.010)\mu+\\
+(-0.631 \pm 0.034)e+(0.586 \pm 0.061)\mu e+\\
+(0.150 \pm 0.041)e^2+(-0.198 \pm 0.047)\mu e^2,
\end{array}
\label{holwie}
\end{equation}
where $\mu = m_B/(m_A+m_B)$ and $a$ and $e$ are the semimajor axis and the eccentricity of the stellar relative orbit, respectively. The approximation is valid in  the range $0.1 \leq \mu \leq 0.9$, for $0.0 \leq e \leq 0.8$, with the error smaller than 11\% (as claimed by the authors). In their approach, the authors assumed that the planet orbits were initially circular, which means that the results obtained using Equation (\ref{holwie}) are valid only along the $y$--axis in Figure \ref{zerovel-all}. For purpose of comparison with our criterion predictions, we re-write the Holman \& Wiegert (1999) criterion  as $a_{p}(1+e_p) < F_{cr}$ and apply it to the $\alpha$ Centauri binary.

The result is shown by the blue curves in Figure \ref{zerovel-all}. The correlation with the MAcD--criterion is notable, but it is clear that the \cite{1999AJ....117..621H} criterion overestimates the instabilities in the planet motion, in this way, excluding from the study a significative portion of still stable configurations. It is worth noting that the Holman \& Wiegert (1999) criterion (\ref{holwie}), also as other empirical criteria (e.g., \citealp{1988A&A...191..385R}, \citealp{2002CeMDA..82..143P}, \citealp{2006PASP..118.1510F}, among others), introduces a concept of \emph{grey area} in the parameter space of the problem, in which the state of a system changes from stability to instability \citep[Chapter 11]{2010ASSL..366.....H}. The grey area is defined as the confidence interval which measures the reliability of the estimation obtained from the fitting procedure. Applying the formula (\ref{holwie}) to the $\alpha$ Centauri system and using the traditional law of error propagation \citep{ku1966}, we obtain the confidence intervals for the stars A and B as $F_{cr}^A = 2.80 \pm 0.64$\,AU and $F_{cr}^B = 2.54 \pm 0.69$\,AU, respectively. The borders of the corresponding grey areas on the ($e_p$,$a_p$)--plans are shown by the cyan-colored curves in Figure \ref{zerovel-all}. However, it must be emphasized here that the grey area idea does not apply to the MAcD--criterion (\ref{SPSeq1}), which is founded on the basic concept of the Hill stability and cannot be interpreted in statistical terms. Due to short time intervals of numerical integrations, the method can still originate effects analogous to `grey areas' (seen as red-colored zones in Figure \ref{zerovel-all}), but these zones rapidly decrease with increasing integration timespans.

Also, it is worth noting here that the purposes of both methods are different.  The purpose of the \cite{1999AJ....117..621H} criterion is  a prompt estimation of the stability of one initially circular configuration of a putative planet in the binary, given the distance of the planet to the central star. Notwithstanding the facility in implementation (it does not require numerical integrations), this method still suffers on some limitations. One of these is due to assumption that the planet starts on zero-eccentricity orbit. Due to very strong gravitational perturbations  produced by the stellar companion on the planetary motion, this assumption can scarcely be correct.

Another limitation is related to the detection of the planets in binary systems and consequent determination of their orbits through the best-fitting procedures.  It seems unlikely that the sets of the osculating orbital elements of the planets provided will be compatible with the circular orbits needed in applying the \cite{1999AJ....117..621H} criterion. Our MAcD--criterion can overcome this limitation, since it is based on the implementation of numerical integrations which use the sets of the orbital parameters of the planets as input.

\subsection{Extension to the 3D}\label{sec1-2}

The MAcD--criterion can be easily extended to the inclined to the reference plane orbits.  Indeed, since the maximum distance of the planet from the central star in space is still its apocentric distance, we can apply the same condition of stability $a_p(1+e_p) < R_{cr}$, where $R_{cr}$ is a radius of the sphere of stability given by Equations (\ref{approximation1}), for the star A, and (\ref{approximation2}), for the star B.

\begin{figure*}
\centering
\includegraphics[width=0.98\columnwidth]{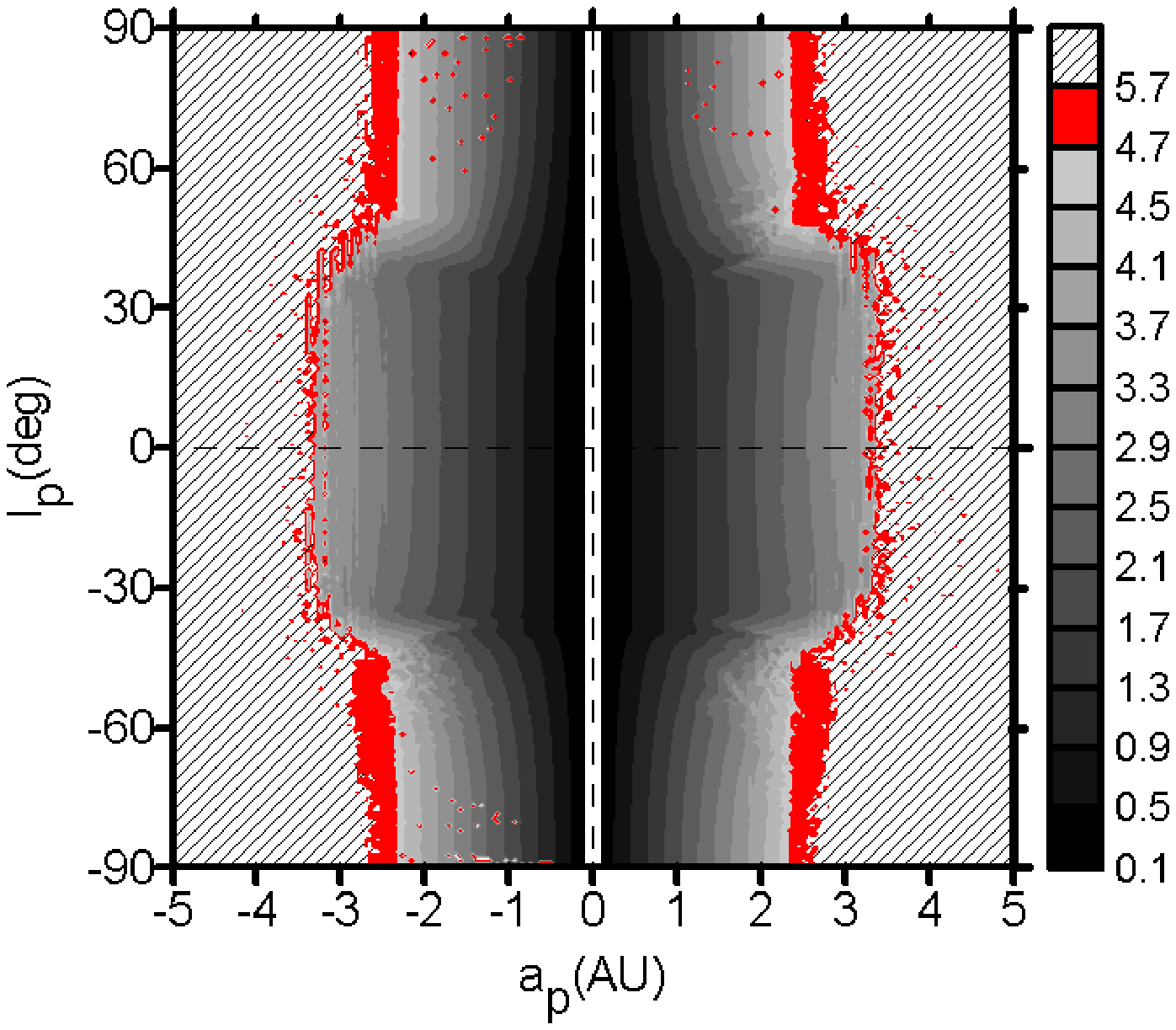}
\hspace{0.5cm}
\includegraphics[width=0.98\columnwidth]{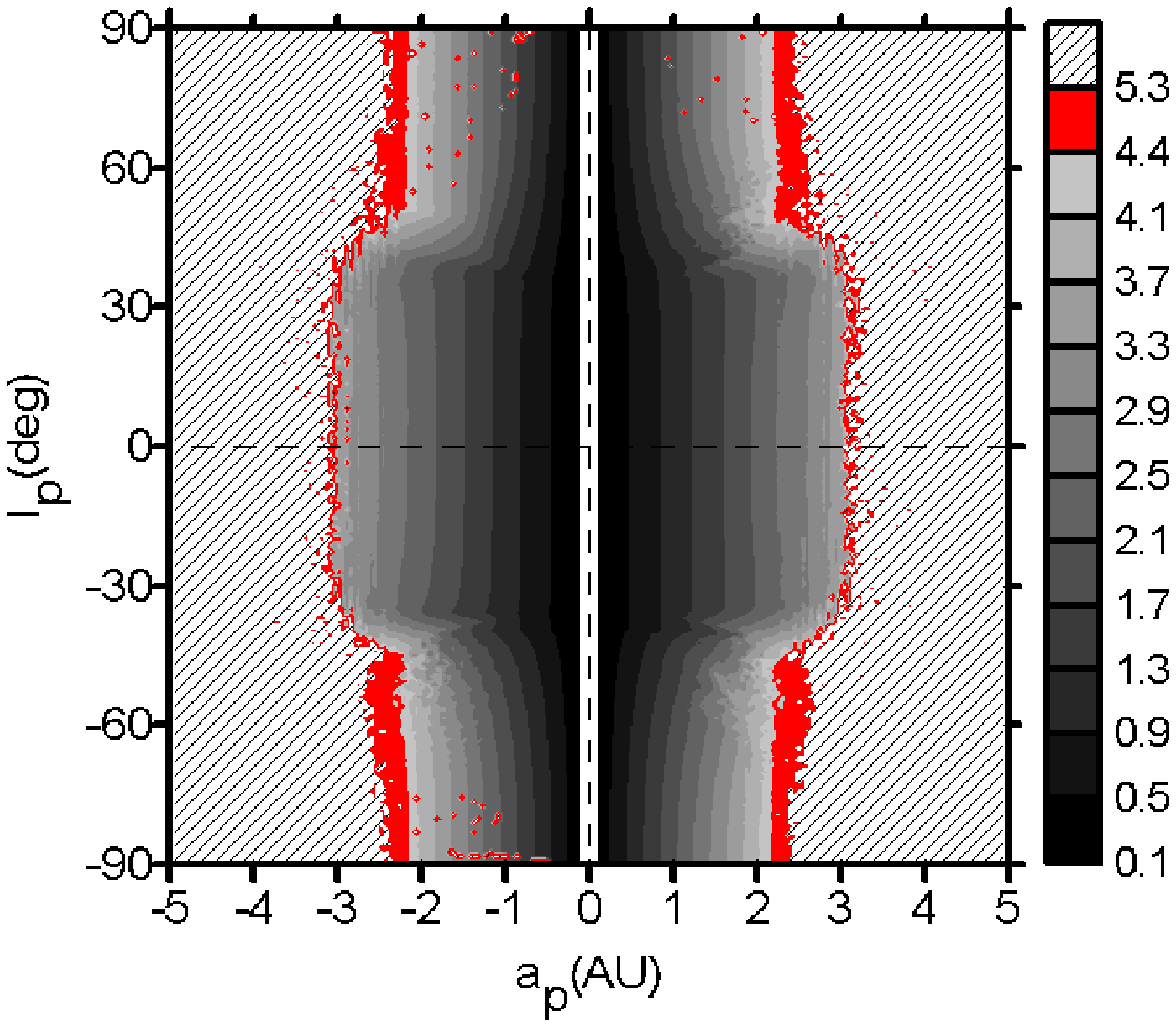}
\caption{Same as in Figure \ref{zerovel-all}, except  in the 3D case, with $a_p$ and $I_p$ being the semimajor axis and the inclination of the planet and the initial eccentricity $e_p$ fixed at $0.05$. The positive and negative values on $x$ ($y$)--axis indicate that the initial $\Delta \varpi$ ($2\omega_p$) are $0^\circ$ or $180^\circ$, respectively. } \label{zerovel-3d}
\end{figure*}

To test the stability of the inclined planetary orbits, we construct the dynamical maps, each with a $400 \times 400$ grid of initial conditions, on the ($a_p$, $I_p$) representative planes, with the planetary eccentricity fixed at $e_p = 0.05$, again with the planetary mass fixed at $m_p = 5\times10^{-5}M_\odot$. Our previous study \citep{2006Icar..181..555M} pointed out that the initial angular conditions of the secular angle $\Delta \varpi$ and the argument of pericenter $2\omega_p$ can be fixed at either $0^\circ$ or $180^\circ$, without loss of generality. The maps obtained are shown in Figure \ref{zerovel-3d}.

We can see that the boundaries of the stable motion are well defined by the MAcD--criterion (red levels), when the apocentric distances approach the critical values $R^A_{{cr}} = 4.7$\,AU and $R^B_{{cr}} =  4.4$\,AU. The maximum values of the planetary semimajor axis, for which the motions still remain stable, are $\approx 3.5$\,AU and $\approx 3$\,AU, for the stars A and B, respectively.  For smaller values of the semimajor axis, there are zones of regular motion for any values of the planetary inclination. The stable motion is robust at inclinations below $\sim 40^\circ$. Above this inclination, we can detect sporadic chaotic solutions (red dots) deeply inside the region of the stable motion. This is an indicator that the analysis of the short-term stability is not sufficient in this case and must be extended to a longer length-times. This will be done in the next section, where we will show that the instabilities at $I_p > 40^\circ$ are originated  by the Lidov--Kozai secular resonance.

The method described in this  section provides a global view of the short-term stability of the system. It is conceptually simple, fast and easily  implementable. It allows us to detect promptly the domains of regular motion, avoiding, in this way,  long-term integrations and analysis of unstable orbits. However, due to the analysis of orbital behaviour of the planets over short timespans, this approach provides no information on the long-term (of order of the lifetime of the stars) stability of the system. The specific study on the stability of the planetary motion in the binary over the age of the system will be done in the next section.

\section{Long-Term Dynamical Stability}\label{sec2}

 Under a ``long-term" stability we mean the stability of the system over the lifetime of the central star. One way to assess the long-term stability of the planet motion is direct N-body numerical simulations, which, depending on the stellar types and ages, must be done over several billions of years. This approach suffers from several limitations and the most serious from these is extremely high computing costs.

An alternative is the dynamical mapping of the domain under study. This approach is also founded on purely numerical integrations, but the time intervals of the simulations are much shorter in this case, and, therefore, we are allowed to explore a large set of initial conditions. Analyzing the topological structure of dynamical maps, we can detect the mechanisms which could induce long-term instabilities in the planetary motion, and determine domains of their action.

The topology analysis is based on the analytical modeling of the general three-body problem briefly described in the following (for more details, see \citealp{2006Icar..181..555M}).  In the Jacobian reference, the Hamiltonian which describes the motion of the planet ($m_p$) around one of the stars ($m_A$), with the second star ($m_B$) as the disturber, is written as \citep{1961mcm..book.....B}

\begin{equation}
{\mathcal H} = {\mathcal H}_{\rm Kep} +  {\mathcal H}_{\rm Pert},
\label{eq:eq0}
\end{equation}

\noindent where:

\begin{equation}
{\mathcal H}_{\rm Kep} = -\frac{G\,m_p\,m_A}{2\,a_p}-\frac{G\,m_B\,(m_A+m_p )}{2\,a_B},
\label{eq:eq0a}
\end{equation}

\begin{equation}
{\mathcal H}_{\rm Pert} = -\frac{G\,m_p\,m_B}{r_{pB}}-G\,m_A\,m_B\left(\frac{1}{r_{AB}}-\frac{1}{r_2}\right),
\label{eq:eq0b}
\end{equation}

\noindent and $a_i$ is the osculating semimajor axis of $i$-th orbit, $\vec{r}_B$ is the position vector of $m_B$ relative to the center of mass of $m_A$ and $m_p$ and $\vec{r}_{iB}$ is the distance between $m_i$ and $m_B$; $G$ is the gravitational constant. Hereafter, the indices $i= A,p,B$ stand for the central star, planet and perturbing star orbits, respectively.

Associated with the Keplerian part of the Hamiltonian, a set of mass-weighted Delaunay's elliptic variables is introduced as
\begin{equation}
\begin{array}{r@{=}lc@{=}l}
M_i & {\rm mean\,\,anomaly,} & L_i &m_i^\prime\sqrt{\mu_ia_i}{\rm ,}\\
\omega_i & {\rm argument\,\,of\,\,pericenter,} & G_i  &L_i\sqrt{1-e_i^2}{\rm ,}\\
\Omega_i& {\rm longitude\,\,of\,node,} & H_i & L_i\sqrt{1-e_i^2}\,\cos\,I_i{\rm ,}
\end{array}
\label{eq:eq2}
\end{equation}

\noindent where $e_i$ and $I_i$ are the eccentricities and inclinations, respectively; $\mu_p=G(m_A+m_p)$, $\mu_B=G(m_A+m_p+m_B)$, $m_p^\prime=m_p\,m_A/(m_A+m_p)$ and $m_B^\prime=m_B\,(m_A+m_p)/(m_A+m_p+m_B)$.

We assume that the planet and the perturbing star are not involved in any mean-motion resonance of low order. Thus we can perform numerically the averaging of the Hamiltonian (\ref{eq:eq0}) with respect to the mean anomalies of the planet and the disturber \citep{2004Icar..168..237M}. The secular Hamiltonian is then defined by
\begin{equation}
{\overline {\mathcal H}_{\rm s}} = -\frac{1}{{(2\pi)^2}}\int_0^{2\pi}\int_0^{2\pi}
R(L_i,G_i,H_i,M_i,\omega_i,\Omega_i)\,dM_pdM_B, \label{eq:eq3}
\end{equation}
where the Keplerian part is constant and therefore needs not be considered.

\begin{figure*}
\centering
\includegraphics[width=0.95\textwidth]{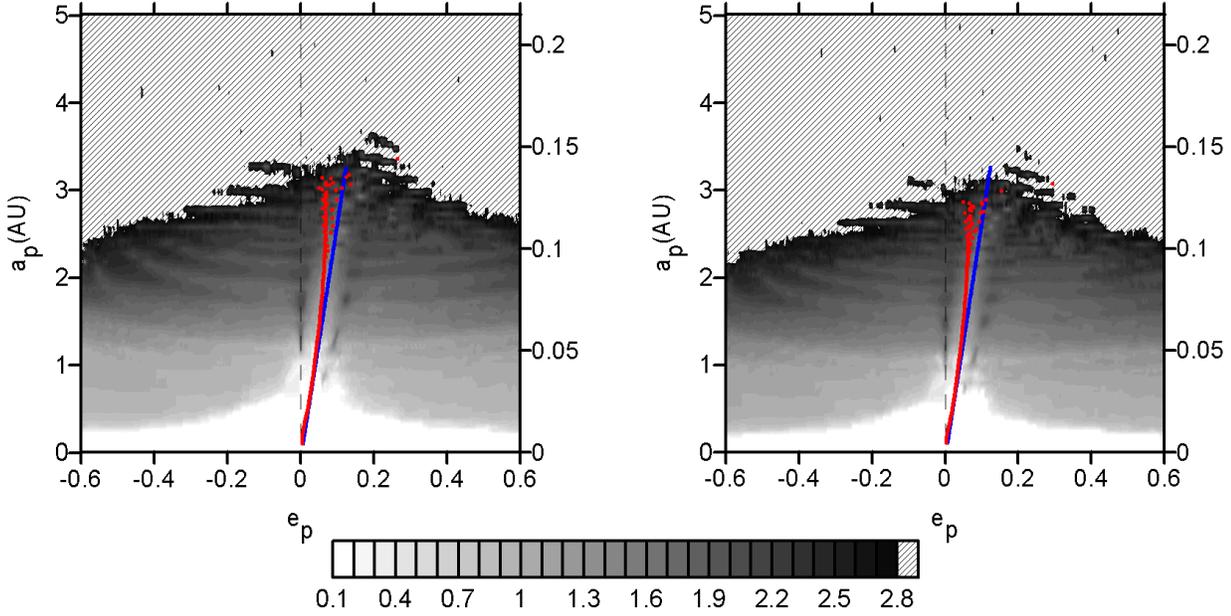}
\caption{Dynamic maps around $\alpha$ Centauri A (left panel) and B (right panel): long-term stability in the planar case. The grey scale code represents the spectral number (stability indicator) in logarithmic scale: the lighter tones correspond to regular motion, while the darker tones correspond to increasingly chaotic motion. The positive and negative values along the $x$-axis indicate that the initial values of $\Delta \varpi$ are $0^\circ$ or $180^\circ$, respectively. The hatched patterns represent ejected or collided orbits. The blue curves are families of stationary solutions provided by the secular semi-analytical Hamiltonian (\ref{eq:eq3}), while the red curves are families obtained through numerical averaging \citep{2011A&A...530A.103G}. The right-hand axes on both planes are the planetary semimajor axis in unit of the semimajor axis of the binary ($a_p/a_B$). } \label{logN-planar}
\end{figure*}

\subsection{Planar case}\label{sec2-1}

In this section, we construct the dynamical maps of coplanar motions, using a grid of $392 \times 100$ osculating initial conditions on the (${e}_p,{a}_p$)-plane and fixing the initial angular variables at $M_p = M_{A,B} = 0$ and $ \Delta\varpi = 0$. Each initial condition was integrated over $10^4$ years ($\sim125$ orbital periods of the binary or $\sim10$ secular periods, see Figure \ref{dinamespec}), with the planetary mass fixed at $m_p=5\times10^{-5}M_\odot$. The solution obtained was Fourier analyzed,  in order to calculate the spectral number corresponding to the used initial condition. The spectral number defines the  degree of chaos in the planetary motion: its small values correspond to regular motions, while its larger values correspond to  unstable motions \citep{2002Icar..158..343M,2005LNP...683..219F}.

The results are shown in Figure \ref{logN-planar}. The grey color scale is used to plot the spectral number: light regions correspond to small numbers and regular motions, while the increasingly dark tones indicate large numbers and unstable orbits. The hatched patterns correspond to the domains where the MAcD--criterion of stability is not satisfied over the chosen integration timespan (see Sect. \ref{sec1-1}). The blue curves show the families of the stable stationary solutions of the secular problem given by the Hamiltonian model (\ref{eq:eq3}), frequently referred to as Mode I of motion \citep{Michtchenko2001}; these solutions are also known as forced eccentricities of the secular problem \citep{1978A&A....65..421H,2011A&A...530A.103G}. In the planar case, the secular problem is reduced to one degree of freedom and the secular solutions are obtained by plotting the energy and angular momentum levels on the representative plane (for a detailed description, see \citealp{2004Icar..168..237M}). The secular planetary motion projected on the (${e}_p,{a}_p$)-plane is an oscillation around the corresponding equilibrium belonging to the blue curve. The closer the initial condition is to the Mode I solution, the smaller is the amplitude of the oscillation of its eccentricity. This fact explains the very stable, nearly harmonic, dynamics of the planet in the vicinity of the blue curve (light grey region), even at larger values of $a_p$.

However, it is worth stressing that the initial conditions used in construction of the dynamical maps are osculating orbital elements; thus, the comparison of the results of numerical integrations to those provided by the secular model (\ref{eq:eq3}) should be done carefully. Moreover, the mass of the disturbing body is comparable to the mass of the central star; this fact reduces the domain of the application of the first-order in masses secular model (\ref{eq:eq3}). We plot in Figure \ref{logN-planar}, by the red curves, the families of the secular stationary solutions obtained through the purely numerical filtering procedure, as described in \cite{2011A&A...530A.103G}. The deviation of the analytically obtained family from that obtained numerically (with no restriction on the perturbing mass) clearly occurs at larger semimajor axes, when the perturbations due to the stellar companion become stronger. Since, in this region, the planetary motion is stable only in the very narrow vicinity of the Mode I, the correct determination of the family is essential in this case.

The two families match each other at small semimajor axes of the planet, up to $\sim 1.5$\,AU (or $a_p/a_B < 0.06$ ), for small eccentricities. The dynamical maps in Figure \ref{logN-planar} show that the planetary dynamics  in this case is very regular (white color), even at very-high eccentricities. It is worth noting that the HZs around the stars are located close to these regions (see Section \ref{sec3}).

\begin{figure*}
\centering
\includegraphics[width=0.95\textwidth]{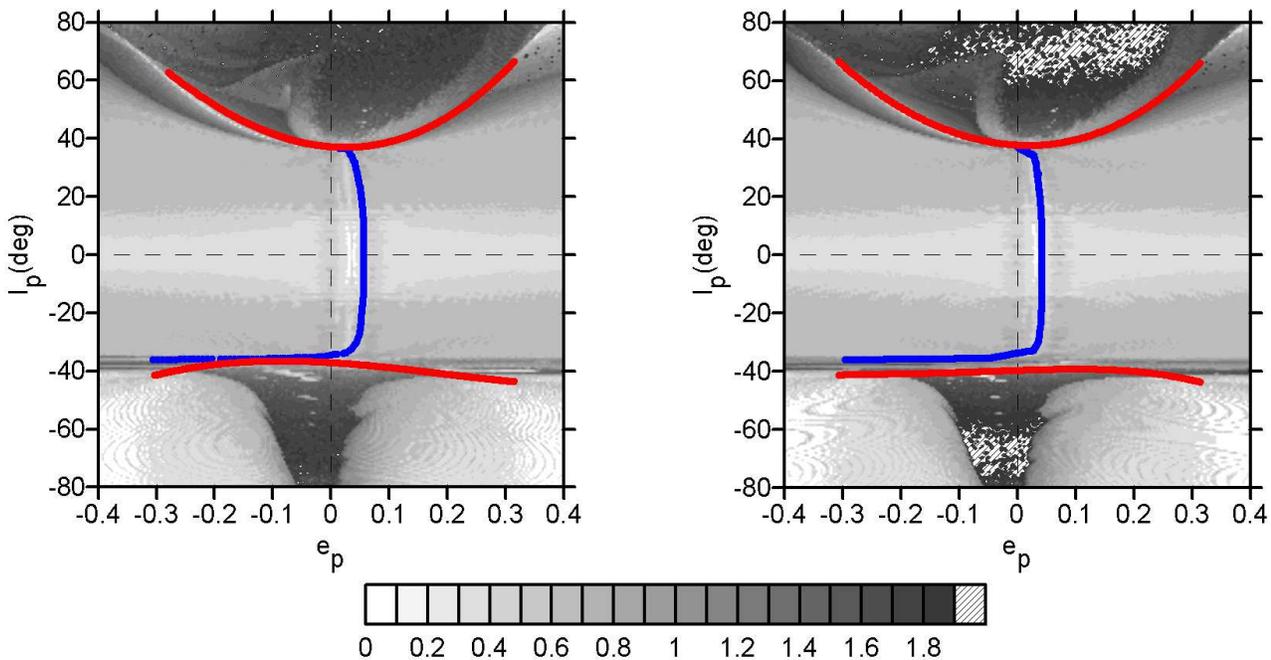}
\caption{Same as in Figure \ref{logN-planar}, except the long-term stability in the 3D case, with $e_p$ and $I_p$ being the eccentricity and the inclination of the planet and the initial semimajor axis fixed at $1.0$\,AU. The positive and negative values on $x$ ($y$)-axis indicate that the initial $\Delta \varpi$ ($2\omega_p$) are $0^\circ$ or $180^\circ$, respectively. The blue and red curves represent the stable and unstable periodic solutions of the Hamiltonian (\ref{eq:eq3}), respectively.}\label{fig:fig4}\end{figure*}

When the semimajor axis of the planetary orbit increases, the perturbations due to the stellar companion become stronger and the planetary motion increasingly non-harmonic  (except in the vicinity of the Mode I of motion). The dark tones on the dynamical maps in Figure \ref{logN-planar} confirm this fact. Several high order mean-motion resonances, from 13/1 to up 20/1, are dense in these regions; they appear as thin horizontal strips on the dynamical maps. The phenomenon can be also observed in the power dynamical spectrum in Figure \ref{dinamespec}, where the resonant initial conditions appear as vertically scattered dots, at $a_p > 2$\,AU. Since these resonances can destabilize the planetary motion,  the long-living planets are not expected to be found in this region.

Finally, at relative distances above 0.14 (in units of the semimajor of the binary orbit), the planetary motion is strongly chaotic and the planets are ejected from the binary (hatched regions). The boundary of these domains match the boundaries of short-term instabilities showed in Figure \ref{zerovel-all}.

It is worth to emphasize the features of the planet motion near the Mode I stationary solutions. Being robust and stable, even at large distances from the central star, this is a favorite location of several known planetary extra-solar systems (e.g., $\upsilon$ Andromedae c-d planet pair). \cite{2011MNRAS.415.2275M} shows that, during migration, the planets are guided toward the stationary configurations, independently on the specific migration mechanism.  Moreover, \cite{2011A&A...530A.103G} shows that the planet formation is preferential on the orbital configurations corresponding to the Mode I stationary solutions.

\subsection{Three-dimensional case}\label{sec2-2}

We construct the dynamical maps on the plane ($e_p$,$I_p$) of the initial osculating values of the planetary eccentricity and the inclination, with a terrestrial planet of mass $m_p=5\times10^{-5} M_\odot$. The spectral analysis method was applied on the  $201\times 201$--grid, with the spacings $\Delta e_p=0.004$ and $\Delta I_p=0.8^\circ$. The initial semimajor axis of the planet was fixed at $a_p = 1.0$\,AU; for this value, the planet is distant from the perturbing star and the first-order Hamiltonian model (\ref{eq:eq3}) can be used to investigate the topology of the dynamical maps.

The maps are shown in Figure \ref{fig:fig4}. The positive and negative values on the $x (y)$--axis stand for the initial values of the angle $\Delta\varpi_p (2\omega_p)$ of $0^\circ$ and $180^\circ$, respectively (see \citealp{2006Icar..181..555M}, for more details). The power dynamical spectrum in Figure \ref{dinamespec} allows us to estimate the secular period  of the planet at $a_p = 1$\,AU as $\sim 10^4$ years. This means that an appropriate timespan for simulations of the secular dynamics must be of order of at least $10^5$ years. The chosen timespan of integration for our dynamical map was 500,000 years ($\sim 6000$ orbital periods of the binary), for each initial condition on the grid; the orbits disrupted during this time are shown by the hatched pattern in Figure \ref{fig:fig4}.  The surviving orbits are shown by the grey scale code, from light tones, for nearly periodic orbits, to darker tones, for increasingly unstable orbits.


\begin{figure*}
\centering
\includegraphics[width=0.95\textwidth]{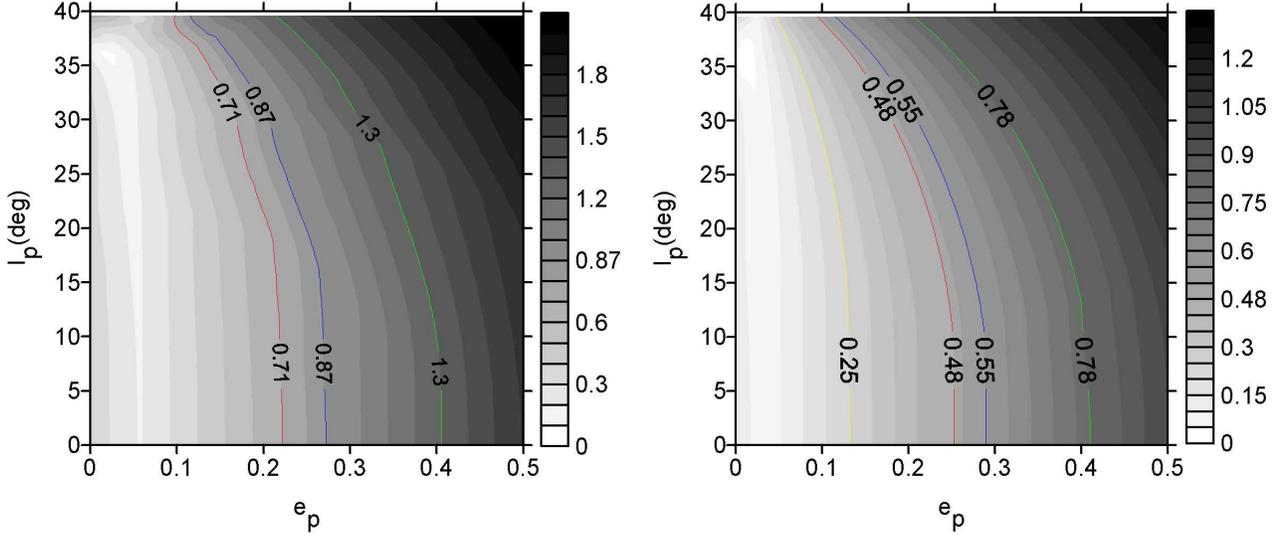}
\caption{Maps of the orbital distances of the planets around the star $\alpha$ Centauri A (left) and the star B (right). The grey scale represents levels of the distance amplitude oscillation of the planet for different initial values of eccentricity and inclinations. The initial semimajor axis is fixed at $a_p = 1.6$\,AU, for $\alpha$ Centauri A, and $a_p = 0.9$\,AU, for $\alpha$ Centauri B. The yellow, red, blue and green curves represent the width of the HZ given by Forgan (2012), Eggl et al. (2013) and Kaltenegger \& Haghighipour (2013), narrow and empirical.} \label{graf2-mapSP}
\end{figure*}

The two color curves represent the periodic solutions obtained through the secular model (\ref{eq:eq3}). The blue curves are location the of the stable periodic solutions, while the red curves are locations of the unstable solution. The long-living motions are expected to be found near the blue curves and far from the red curves. Indeed, the analysis of the structure of the dynamical maps of the representative planes reveals that, for low inclinations $I_p<40^\circ$, the maps are dominated by a light grey background, which means that the low-inclination secular dynamics is regular.

In contrast, the vicinity of the red curves is dominated by the darker tones, indicating unstable motions. The instability is related to the existence of separatrices between different regimes of motion. Their domains, at high inclinations, show complex dynamical behaviour, with the presence of several regimes of resonant motion (the detailed description of these regimes of secular motion can be found in \citealp{2006Icar..181..555M}). The dominating behaviour is the $e_p$--$I_p$ coupling, or Lidov--Kozai resonance, which destabilizes the nearly circular orbits at $I_p>40^\circ$. For higher values of eccentricity, the planetary dynamics is characterized by the coupled large variation of the eccentricity and inclination and the libration of the angle $\omega_p$ around $\pm 90^\circ$. The domains of the Lidov-Kozai resonance are located on the lower half-planes in Figure \ref{fig:fig4}; the behaviour of the system is very regular in these regions.

The $\Delta\varpi$ secular resonance also exists in the high-inclination regions; its domains are thin white strips along the red curves. In this case, the secular angle $\Delta\varpi$ librates either around $0$ or $180^\circ$.

It is worth emphasizing that we have detected a large portion of the chaotic orbits (black tones on the graphs in Figure \ref{fig:fig4}), but only a small quantity of these orbits are actually ejected from the system (dashed regions). This is because the chaotic processes are slow in the secular system and the timespan of 500,000 years is not sufficient for detecting the planetary escapes. However, the escapes are imminent and, as a consequence, it will be impossible to detect the planets inside these regions of high instabilities.

\section{Dynamical evolution inside the HZ}\label{sec3}

Regardless of the robust long-term stability of the planetary motion in the large domains around the stars of $\alpha$ Centauri, the planetary evolution inside the HZs, should be analyzed, in order to verify whether the planet motion is confined inside the HZ. The concept of the HZ itself, even when introduced for single stars (e.g., \citealp{1993Icar..101..108K}), is very complex. In binary stars, the definition of the HZ is further complicated due to several additional aspects. One is the combined radiation of both stellar components on the terrestrial planet studied recently in several papers  (e.g., \citealp{2012MNRAS.422.1241F}, \citealp{2013ApJ...777..165K}).

Another aspect concerns the large oscillations of the orbital elements of the planet, that could affect the conditions of habitability of the planet or even drive the planet far from the HZ. This behaviour was studied in \cite{2012ApJ...752...74E,2013MNRAS.428.3104E}, where the authors considered the averaged effects produced by the secondary star on the variation of the eccentricity of the planet, yet in the case of coplanar and initially circular planetary orbits. In this work we explore the conditions for which the planets remain inside the HZ, in the case of eccentric and inclined orbits. For this, we use the different definitions of the HZs around the $\alpha$ Centauri stars given in the papers \cite{2012MNRAS.422.1241F}, \cite{2013MNRAS.428.3104E} and \cite{2013ApJ...777..165K}(see Table \ref{tableHZ}). It should be emphasized that these definitions are not equally weighted. In fact, the KH13 definition can be considered more realistic because it takes into account the presence of an atmosphere on a terrestrial or super-Earth planet. However, the detailed discussion on the HZ calculations is out of the scope of this paper.

For each definition from Table \ref{tableHZ}, we calculate the maximum and minimum distances of the planet from the central star, for different initial eccentricities and inclinations of the planet orbit and the initial semimajor axis fixed at $a_p = 1.6$\,AU, around $\alpha$ Centauri A, and $a_p = 0.9$\,AU, around $\alpha$ Centauri B; these values roughly correspond to the averaged positions of all HZs defined in Table \ref{tableHZ}.

\begin{table}
\caption{
The inner and outer boundaries of the HZs (in AU) and the HZs width in the $\alpha$ Centauri AB binary, calculated in Forgan (2012),  Eggl et al.(2013) and Kaltenegger \& Haghighipour (2013) (KH13) (conservative and optimistic definitions).
}
\begin{center}
\begin{tabular}{lcccc}
Authors& Star & Inner HZ & Outer HZ & Width\\
 \hline
\cite{2013MNRAS.428.3104E} & A & 1.12 & 1.81& 0.71 \\
KH13 conservative & A & 1.20 & 2.07 & 0.87\\
KH13 optimistic & A& 0.92 & 2.18 & 1.30\\
\hline
\cite{2012MNRAS.422.1241F} & B & 0.65 & 0.90 & 0.25\\
\cite{2013MNRAS.428.3104E} & B & 0.65 & 1.13 & 0.48\\
KH13 conservative & B & 0.71 & 1.26 & 0.55\\
KH13 optimistic & B& 0.54 & 1.32 & 0.78\\
\end{tabular}
\label{tableHZ}
\end{center}
\end{table}

The numerical test is done using numerical integrations of planetary orbits over 10,000 years. The values of the maximal and minimal distances obtained are compared to the width of the HZs, in order to infer whether the planet remain within the HZ boundaries over 10,000 years ($\sim 125$ orbital periods of the binary).

The difference between the maximal and minimal distances of the planet from the central star is plotted on the ($e_p$, $I_p$)--planes in Figure \ref{graf2-mapSP}, using the grey scale code: the difference is smaller in the light regions and larger in the dark regions. The regions in white tone correspond to the minimal possible deviations of the planet from the center of the HZ. The simple comparison with the graphs in Figure \ref{fig:fig4} allows us conclude that these regions are associated with the Mode I secular solutions whose location is given by blue curves in Figure \ref{fig:fig4}.

An interesting feature of this analysis is that initially circular orbits exhibit significant deviations from the center of the HZ, that, for high inclinations, could result in the planets spending some time outside the HZs.

As discussed in \cite{2012ApJ...752...74E}, the planet remaining inside the HZ during its evolution is not the only requirement for the planet to host life, and it is neither an excluding factor. Even if the planet spends some time of its orbit outside the HZ, as long as the mean radiation received is still within the HZ limits, the planet could be considered habitable, although in a more restrict way. Nevertheless, it is expected that orbits with smaller oscillations are able more likely to sustain some kind of life.

\begin{figure}
\centering
\includegraphics[angle=270,width=1.0\columnwidth]{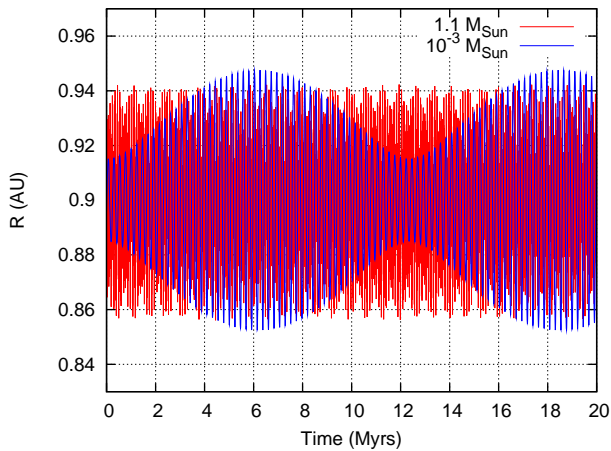}
\caption{Time evolution of the planetary distance to $\alpha$ Centauri B when the perturber is $\alpha$ Centauri A (red) and a Jupiter-mass planet (blue).}
\label{orbits}
\end{figure}

 It is also interesting to compare the orbital evolution of a terrestrial S-type planet in $\alpha$ Centauri AB  with the evolution of the terrestrial planet in a planetary system, with a Jupiter-sized planet. The comparison is done using the configurations defined  by the orbital parameters of $\alpha$ Centauri AB, except, in the case of the planetary system, the binary companion is replaced by a Jupiter-mass planet. The results are  shown in Figure \ref{orbits}, where the variation of the orbital distance of the planet revolving around $\alpha$ Centauri B is plotted by the red curve, in the case when the perturber is the star $\alpha$ Centauri A ($1.1 M_\odot$), and by the black curve, when the perturber is a hypothetical Jupiter-mass planet ($10^{-3}M_\odot$). In both simulations, the same orbital and physical parameters of the planet and perturbing body were used, except the mass of the perturber. The initial semimajor axis and the eccentricity of the planet were fixed at $a_p = 0.9$\,AU (inside the HZ of the star) and $e_p = 0.016$, respectively; note that this value of the eccentricity corresponds to the actual eccentricity of Earth. The initial orbital elements of the perturber were the same of the relative orbit of the $\alpha$ Centauri binary.

We can observe in Figure \ref{orbits} that the amplitudes of the oscillation of the orbital distance of the planet are of the same order of magnitude, despite that the perturbing masses differ in order 3. This result is consistent with the theoretical predictions of the restricted three-body models (e.g., \citealp{1978A&A....65..421H}) which state that the forced eccentricity of the test particle is independent on the masses of the large bodies (at least, in first order in masses). Therefore, the orbital distance oscillation of the planet in the binary system is  similar to that of a terrestrial planet perturbed by a giant planet. The main difference is the timescale of the oscillation: it is of order of $10^{4}$ years, in the binary system, and of order of $10^7$ years, in the planetary system. However, this difference seems to be not important in the context of the Continuously Habitable Zone.

In addition, we have obtained that, in the case of the real Earth-Jupiter system, the forced eccentricity of the Earth's orbit is equal to $0.016$; in the case of a putative terrestrial planet orbiting $\alpha$ Centauri B,  the forced eccentricity of the planetary orbit is equal to $0.032$. Calculating the variation of the orbital distances in both cases, we have found that its amplitude is $\sim 0.032$\,AU, for the Earth's orbit, and $\sim 0.07$\,AU, for the planet located near the Mode I stationary solution of the HZ of $\alpha$ Centauri B, that is, both variations are of same order of amplitude. This result brings positive perspectives of finding habitable planets in binary star systems.

\section{Summary}\label{sum}

In this work, we present a methodology to analyze the planetary dynamics of S-type orbits in binary star systems. We develop a short-term stability criterion (MAcD--criterion), which is based on the Hill concept of stability. Since this method does not require any kind of expansion and averaging of the disturbing function, there are no constrains on neither the mass of the disturbing body nor the eccentricity/inclination of its orbit. We apply our approach to the $\alpha$ Centauri binary system with a hypothetical Earth-like planet, orbiting one of the stars (S-type orbit), and show that the purposed criterion provides a global view of the stable and unstable domains in the space of the orbital elements. It is worth emphasizing that our approach is simple, fast and easily implementable. Its main purpose is to detect promptly the domains of regular motion, avoiding, in this way,  long-term integrations and analysis of unstable orbits. It also allows us to quickly assess the stability of the best-fit solutions in the determination of the planetary orbits hosted by binary stars.

The long-term stability of the planetary orbits in the HZs of the $\alpha$ Centauri binary system is studied using the secular Hamiltonian approach. We show that the behaviour of the planets is well described by the first-order in masses model. The secular planetary motion is regular, even at high eccentricity domains of the HZs.  There are no significant mean-motion resonances that could destabilize the secular dynamics. The stability of the planets is also detected through the 3D--analysis,  at low and moderate inclinations of the planetary orbits ($ < 40^\circ$). For higher inclinations,  the dominating regime of motion is the Lidov--Kozai resonance, which is separated from the purely secular regime of motion through the layers of strongly chaotic motion. Even though there are domains of very stable motion in the Lidov--Kozai resonance, the large excursions of the planetary eccentricities and inclinations seem to be unfavorable for the evolution of life inside the HZs. Beyond the HZ, where high-order mean-motion resonances take place, the first-order model fails. Thus, a second-order in masses model, as well as the modeling of the resonant behaviour, would be required to further study of this region.

 Analyzing  the dynamics of the planets evolving inside the HZs of the $\alpha$ Centaury stars, we obtain that the planetary orbits will remain within the boundaries of the HZs, if their eccentricities and inclinations are constrained to values close to the stationary solutions of the secular problem, the Mode I solutions. In this favorable scenario, the amplitude of the orbital distance variations are comparable to the amplitudes of the Earth's orbit, with the only difference being the timescales of the variations. The real effects of this on the planetary climate and geological conditions are still poorly known and it is expected that further interdisciplinary studies could bring to light considerable new information on this issue.

\section{Acknowledgments}

This work was supported by the S\~ao Paulo State Science Foundation, FAPESP (grant 2010/01209-2), and the Brazilian National Research Council, CNPq. This work has made use of the facilities of the Computation Center of the University of S\~ao Paulo (LCCA-USP) and of the Laboratory of Astroinformatics (IAG/USP, NAT/Unicsul), whose purchase was made possible by the Brazilian agency FAPESP (grant 2009/54006-4) and the INCT-A.  The authors are grateful to Dr. Haghighipour for numerous helpful suggestions/corrections on this paper.


\begin{thebibliography}{}


\bibitem[Abt(1979)]{1979AJ.....84.1591A} Abt, H.~A.\ 1979, Astronomical Journal, 84, 1591

\bibitem[Barbieri et
al.(2002)]{2002A&A...396..219B} Barbieri, M., Marzari, F., \& Scholl, H.\ 2002, Astronomy and Astrophysics, 396, 219

\bibitem[Black(1982)]{1982AJ.....87.1333B} Black, D.~C.\ 1982, Astronomical Journal, 87, 1333

\bibitem[Boss(2006)]{2006ApJ...641.1148B} Boss, A.~P.\ 2006, Astrophysical Journal, 641, 1148

\bibitem[Brouwer \& Clemence(1961)]{1961mcm..book.....B} Brouwer, D., \& Clemence, G.~M.\ 1961, New York: Academic Press, 1961,

\bibitem[Dumusque et al.(2012)]{2012Natur.491..207D} Dumusque, X., Pepe, F., Lovis, C., et al.\ 2012, Nature, 491, 207

\bibitem[Duquennoy \& Mayor(1991)]{1991A&A...248..485D} Duquennoy, A., \& Mayor, M.\ 1991, Astronomy and Astrophysics , 248, 485

\bibitem[Dvorak et al.(1986)]{1986BAAS...18..842D} Dvorak, R., Froeschle, C., \& Froeschle, C.\ 1986, Bulletin of the American Astronomical Society, 18, 842

\bibitem[Eggl et al.(2013)]{2013MNRAS.428.3104E} Eggl, S., Pilat-Lohinger, E., Funk, B., Georgakarakos, N., \& Haghighipour, N.\ 2013, Monthly Notices of the Royal Astronomical Society, 428, 3104

\bibitem[Eggl et al.(2012)]{2012ApJ...752...74E} Eggl, S., Pilat-Lohinger, E., Georgakarakos, N., Gyergyovits, M., \& Funk, B.\ 2012, Astrophysical Journal, 752, 74

\bibitem[Everhart(1985)]{1985dcto.proc..185E} Everhart, E.\ 1985, Dynamics of Comets: Their Origin and Evolution, Proceedings of IAU Colloq.~83, held in Rome, Italy, June 11-15, 1984.~ Edited by Andrea Carusi and Giovanni B.~Valsecchi.~ Dordrecht: Reidel, Astrophysics and Space Science Library.~Volume 115, 1985, p.185, 185

\bibitem[Fatuzzo et al.(2006)]{2006PASP..118.1510F} Fatuzzo, M., Adams,
F.~C., Gauvin, R., \& Proszkow, E.~M.\ 2006, PASP, 118, 1510

\bibitem[Ferraz-Mello et al.(2005)]{2005LNP...683..219F} Ferraz-Mello, S.,
Michtchenko, T.~A., Beaug{\'e}, C., \& Callegari, N.\ 2005, Chaos and Stability in Planetary Systems, 683, 219

\bibitem[Ford et al.(2000)]{2000ApJ...535..385F} Ford, E.~B., Kozinsky, B., \& Rasio, F.~A.\ 2000, Astrophysical Journal, 535, 385

\bibitem[Forgan (2012)]{2012MNRAS.422.1241F} Forgan, D.\ 2012, Monthly Notices of the Royal Astronomical Society, 422, 1241

\bibitem[Giuppone et al.(2011)]{2011A&A...530A.103G} Giuppone, C.~A., Leiva, A.~M., Correa-Otto, J., \& Beaug{\'e}, C.\ 2011, Astronomy and Astrophysics , 530, A103

\bibitem[Gladman (1993)]{1993Icar..106..247G} Gladman, B.\ 1993, Icarus, 106, 247

\bibitem[Graziani \& Black(1981)]{1981ApJ...251..337G} Graziani, F., \& Black, D.~C.\ 1981, Astrophysical Journal, 251, 337

\bibitem[Guedes et al.(2008)]{2008ApJ...679.1582G} Guedes, J.~M., Rivera,
E.~J., Davis, E., et al.\ 2008, Astrophysical Journal, 679, 1582

\bibitem[Haghighipour(2006)]{2006ApJ...644..543H} Haghighipour, N.\ 2006, Astrophysical Journal, 644, 543

\bibitem[Haghighipour(2010)]{2010ASSL..366.....H} Haghighipour, N.\ 2010,
Astrophysics and Space Science Library, 366

\bibitem[Haghighipour et al.(2010)]{2010ASSL..366..285H} Haghighipour, N., Dvorak, R., \& Pilat-Lohinger, E.\ 2010, Astrophysics and Space Science Library, 366, 285

\bibitem[Haghighipour \& Raymond(2007)]{2007ApJ...666..436H} Haghighipour, N., \& Raymond, S.~N.\ 2007, Astrophysical Journal, 666, 436

\bibitem[Hatzes(2013)]{2013ApJ...770..133H} Hatzes, A.~P.\ 2013, Astrophysical Journal, 770,
133


\bibitem[Heppenheimer(1978)]{1978A&A....65..421H} Heppenheimer, T.~A.\ 1978, Astronomy and Astrophysics , 65, 421

\bibitem[Hill(1878)]{1905Hill}Hill, G. W., 1878, Am. J. Math. 1-5, 129

\bibitem[Holman \& Wiegert(1999)]{1999AJ....117..621H} Holman, M.~J., \& Wiegert, P.~A.\ 1999, Astronomical Journal, 117, 621

\bibitem[Kaltenegger \& Haghighipour(2013)]{2013ApJ...777..165K} Kaltenegger, L., \& Haghighipour, N.\ 2013, Astrophysical Journal, 777, 165

\bibitem[Kasting et al.(1993)]{1993Icar..101..108K} Kasting, J.~F.,
Whitmire, D.~P., \& Reynolds, R.~T.\ 1993, Icarus, 101, 108

\bibitem[Kervella et al.(2003)]{2003A&A...404.1087K} Kervella, P., Th{\'e}venin, F., S{\'e}gransan, D., et al.\ 2003, Astronomy and Astrophysics , 404, 1087

\bibitem[Ku(1966)]{ku1966}Ku, H. H., Journal of Research of the National Bureau of Standards. Section C: Engineering and Instrumentation, 70C, No. 4, 263

\bibitem[Laskar \& Bou{\'e}(2010)]{2010A&A...522A..60L} Laskar, J., \& Bou{\'e}, G.\ 2010, Astronomy and Astrophysics , 522, A60

\bibitem[Marchal \& Bozis(1982)]{1982CeMec..26..311M} Marchal, C., \& Bozis, G.\ 1982, Celestial Mechanics, 26, 311

\bibitem[Michtchenko \& Ferraz-Mello(2001)]{Michtchenko2001} Michtchenko, T.~A., \& Ferraz-Mello, S.\ 2001, Icarus, 149, 357

\bibitem[Michtchenko et al.(2002)]{2002Icar..158..343M} Michtchenko, T.~A.,
Lazzaro, D., Ferraz-Mello, S., \& Roig, F.\ 2002, Icarus, 158, 343

\bibitem[Michtchenko et al.(2006)]{2006Icar..181..555M} Michtchenko, T.~A., Ferraz-Mello, S., \& Beaug{\'e}, C.\ 2006, Icarus, 181, 555

\bibitem[Michtchenko \& Malhotra(2004)]{2004Icar..168..237M} Michtchenko, T.~A., \& Malhotra, R.\ 2004, Icarus, 168, 237

\bibitem[Michtchenko \& Rodr{\'{\i}}guez(2011)]{2011MNRAS.415.2275M} Michtchenko, T.~A., \& Rodr{\'{\i}}guez, A.\ 2011, Monthly Notices of the Royal Astronomical Society, 415, 2275

\bibitem[Nelson(2000)]{2000ApJ...537L..65N} Nelson, A.~F.\ 2000, Astrophysical Journal, 537, L65

\bibitem[Pendleton \& Black(1983)]{1983AJ.....88.1415P} Pendleton, Y.~J., \& Black, D.~C.\ 1983, Astronomical Journal, 88, 1415

\bibitem[Pilat-Lohinger
\& Dvorak(2002)]{2002CeMDA..82..143P} Pilat-Lohinger, E., \& Dvorak, R.\ 2002, Celestial Mechanics and Dynamical Astronomy, 82, 143


\bibitem[Pourbaix et
al.(2002)]{2002A&A...386..280P} Pourbaix, D., Nidever, D., McCarthy, C., et al.\ 2002, Astronomy and Astrophysics, 386, 280

\bibitem[Quintana et al.(2002)]{2002ApJ...576..982Q} Quintana, E.~V.,
Lissauer, J.~J., Chambers, J.~E., \& Duncan, M.~J.\ 2002, Astrophysical Journal, 576, 982

\bibitem[Quintana et al.(2007)]{2007ApJ...660..807Q} Quintana, E.~V.,
Adams, F.~C., Lissauer, J.~J., \& Chambers, J.~E.\ 2007, Astrophysical Journal, 660, 807

\bibitem[Rabl
\& Dvorak(1988)]{1988A&A...191..385R} Rabl, G., \& Dvorak, R.\ 1988, Astronomy and Astrophysics, 191, 385

\bibitem[Raghavan et al.(2010)]{2010ApJS..190....1R} Raghavan, D., McAlister, H.~A., Henry, T.~J., et al.\ 2010, Astrophysical Journal Supplement, 190, 1

\bibitem[Szebehely(1984)]{1984CeMec..34...49S} Szebehely, V.\ 1984, Celestial Mechanics, 34, 49

\bibitem[Th{\'e}bault et al.(2008)]{2008MNRAS.388.1528T} Th{\'e}bault, P.,
Marzari, F., \& Scholl, H.\ 2008, Monthly Notices of the Royal Astronomical Society, 388, 1528

\bibitem[Th{\'e}bault et al.(2009)]{2009MNRAS.393L..21T} Th{\'e}bault, P., Marzari, F., \& Scholl, H.\ 2009, Monthly Notices of the Royal Astronomical Society, 393, L21

\end{thebibliography}
\end{document}